\documentclass[prl,twocolumn,showpacs,preprintnumbers,
               superscriptaddress]{revtex4}
\usepackage{graphicx,epsf}

\begin{document}
\draft

\title{Linear Langevin Equation of Critical Fluctuation in 
Chiral Phase Transition}

\author{T.~Koide
%\footnote{koide@th.physik.uni-frankfurt.de}
}
\affiliation{Institute f\"ur Theoretische Physik, J.~W.~ Goethe-Universit\"at, 
D-60054 Frankfurt, Germany}
\author{M.~Maruyama
%\footnote{maruyama@nucl.phys.tohoku.ac.jp}
}
\affiliation{Department of Physics, Tohoku University, Sendai, Japan}

\begin{abstract}
We derive the linear Langevin equation 
that describes the behavior of critical fluctuation above the 
critical temperature of the chiral phase transition 
in the Nambu-Jona-Lasinio model.
The Langevin equation relaxes exhibiting oscillation and 
shows thermalization.
The relaxation becomes slower as the system approaches 
the critical point(the critical slowing down).
The time correlation function also calculated using the Langevin equation 
shows that there exists the same soft mode as the previous calculation 
in the linear response theory.
\end{abstract}

\pacs{05.10.Gq,11.10.Wx,11.30.Rd}
\date{\today}
\maketitle

It is widely believed that 
hadronic matters undergoes a phase transition at high temperature, 
and hence becomes the quark-gluon plasma (QGP) where 
quarks and gluons are deconfined and chiral symmetry broken in 
the hadron phase is restored.
In order to explore the QGP, 
relativistic heavy-ion collisions are a significant experimental instrument.
Heavy-ion collisions are essentially nonequilibrium processes and thus 
the description of the time evolution is inevitable to understand 
the phenomena in a comprehensive way.
One of the powerful tools to describe time dependent processes 
is the hydrodynamic model.
In some instances, the hydrodynamic model fairly correctly describes 
the experimental data of heavy-ion collisions, for example, 
transverse momentum dependence of the elliptic flow coefficient $v_{2}$
\cite{ref:EF}.

Recently, our attention has been focused 
on the dynamics near the critical point 
of the chiral phase transition\cite{ref:Raja}.
Unfortunately, the hydrodynamic model is sometimes inadequate to describe 
phenomena near the critical point.
The hydrodynamic model is valid only in the case where 
we are interested in gross variables 
associated with macroscopic time and length scales 
(entropy density, energy density, etc) 
and 
they are widely separated from microscopic variables 
(degrees of freedom of each quark and gluon).
However, near the critical point of the second order phase transition, 
the indefinite increase of the correlation range of 
critical fluctuations destroys 
the simple picture of widely separated scales.
This situation is similar to the case of the classical dilute gas, where 
the hydrodynamic model loses its validity in describing 
physics with mesoscopic scales (for example, mean free path), 
and a kinetic equation becomes appropriate.
To describe the critical dynamics, 
it is probable that 
we still do not need the full details of the 
Heisenberg equation of motion because of the universality of phase transition.
Then, from an analogy to gas dynamics, 
the critical dynamics will be described by 
a kinetic equation, like a Langevin equation\cite{ref:Kawasaki}.
The purpose of the present paper is to derive a linear Langevin equation 
to describe the dynamics of the critical fluctuation above the 
critical temperature of the chiral phase transition at zero density.

As a low-energy effective model of QCD, 
we adopt the two-flavor and three-color 
Nambu-Jona-Lasinio (NJL) model\cite{ref:HK-PR};
\begin{eqnarray}
H &=& 
     \int d^3{\bf x} \bar{q}(-i\vec{\gamma}\cdot \nabla)q 
-g \int d^3 {\bf x} \{
          (\bar{q}q)^2
          +(\bar{q}i\gamma_{5}\tau )^2
          \}\nonumber \\
&\equiv& H_{0} + H_{I}.
\end{eqnarray}
The coupling constant $g$ and the three momentum cutoff $\Lambda$ are 
fixed so as to reproduce 
the pion decay constant $f_{\pi} = 93~{\rm MeV}$ and 
the chiral condensate $\langle \bar{q}q \rangle = (-250~{\rm MeV})^3$:
$g = 5.01~{\rm GeV^{-2}}$ and $\Lambda = 650~{\rm MeV}$.
The critical temperature of 
the chiral phase transition using this parameter set 
is $T_{c} = 185$ MeV in the mean field approximation.

In principle, the dynamics of the chiral phase transition is 
described by solving the Heisenberg equation of motion for the order 
parameter, that is, 
the chiral condensate $\langle \sigma \rangle = \langle \bar{q}q \rangle$.
However, the Heisenberg equation incorporates 
motion on microscopic and macroscopic scales on an equal footing.
To carry out coarse-graining of microscopic variables systematically, 
we introduce projection operators $P$ and $Q = 1-P$
\cite{ref:NZ,ref:Mori,ref:SH,ref:KM,ref:review}.
This technique was originally introduced by Nakajima\cite{ref:NZ} 
to derive a master equation.
The application to a Langevin equation was attempted 
by Mori\cite{ref:Mori}.
Recently, the unification and generalization of both treatments 
has been realized\cite{ref:SH,ref:KM}.
It should be noted that the elimination of microscopic variables 
does not mean neglecting their effect.
They give rise to dissipation 
terms and noise terms in the coarse-grained macroscopic equations.

Then, the time evolution of the chiral phase transition is described by
\cite{ref:KM} 
\begin{eqnarray}
\frac{d}{dt}\delta\sigma({\bf x},t) 
&=& e^{iLt}PiL\delta\sigma({\bf x},0) \nonumber \\
&&\hspace*{-1.6cm}+ \int^{t}_{0}ds e^{iL(t-s)}
PiLQe^{iQL_{0}Qs}iL\delta\sigma({\bf x},0) 
+
\xi({\bf x},t), \label{eqn:TC-2}
\end{eqnarray}
where $L$ and $L_{0}$ are Liouville operators of 
the Hamiltonian $H$ and $H_{0}$, respectively.
Here, we calculate the fluctuation of the order parameter 
from its equilibrium value:
$\delta\sigma({\bf x},t) \equiv \sigma({\bf x},t) - <\sigma({\bf x})>_{eq}$.
This equation is called the time-convolution (TC) equation\cite{ref:SH,ref:KM}.
The first term on the r.h.s of the equation expresses the term 
corresponding to a collective oscillation such as plasma wave, spin wave, etc.
The second term is the memory term that causes dissipation.
Here, we have already expanded the memory term up to lowest order 
in $H_{I}$, following Ref.~\cite{ref:KM}.
The third term is the noise term and we do not give its concrete form here.
This is defined through the fluctuation-dissipation theorem 
of second kind (2nd F-D theorem) as we will see later.

There are several possible projection operators that extract 
slowly varying parts from an operator
\cite{ref:SH,ref:KM,ref:review,ref:KMT-L,ref:TDPO}.
In this paper, we adopt the Mori projection operator(MPO) 
that projects any operators onto the space spanned by gross variables
\cite{ref:Mori}.
To define it, we must realize the time scale of gross variables 
are extremely long compared to that of other microscopic variables.
There are three candidates for gross variables\cite{ref:Kawasaki}:
(i) order parameters, 
(ii) density variables associated with conserved quantities and 
(iii) their products.
However, near critical points of the second order phase transition, 
the order parameters are probably much 
slower than the density variables because of critical slowing down(CSD).
In this sense, we can ignore the time evolution of density variables and 
exclude (ii) from our gross variables.
Furthermore, we ignore (iii) for simplicity.
We shall discuss the importance of (iii) at the end of this paper.
After all, the gross variable relevant in our calculation is 
the order parameter of the chiral phase transition $\sigma({\bf x},t)$.
Then, the MPO in our calculation is defined by
\\
\begin{eqnarray}
\lefteqn{P O} && \nonumber \\ 
&&\hspace{-0.4cm} = \int d^3 {\bf x}d^3 {\bf x}'
(O, \delta \sigma ({\bf x})) \cdot 
(\delta \sigma ({\bf x}), \delta \sigma ({\bf x}'))^{-1} \cdot 
\delta \sigma({\bf x}'), \label{eqn:MPO}
\end{eqnarray}
where $O$ is an arbitrary operator.
The inner product means the canonical correlation,
%\cite{ref:Cano-Corr}
\begin{eqnarray}
({\bf X},{\bf Y})
= \int^{\beta}_{0}\frac{d\lambda}{\beta}
{\rm Tr}[\rho e^{\lambda H_{0}}{\bf X}e^{-\lambda H_{0}}{\bf Y} ] ,
\end{eqnarray}
where $\rho=\exp (-\beta H_{0})/{\rm Tr}[\exp (-\beta H_{0})]$ 
and $\beta = 1/T$.
The system attains the thermal equilibrium state with this temperature $T$.
In the classical limit, the canonical correlation coincides with 
the classical correlation function ${\rm Tr}~[\rho {\bf XY}]$.

Substituting the MPO into the TC equation (\ref{eqn:TC-2}), 
we have 
\begin{eqnarray}
\frac{d}{dt} \delta \sigma({\bf k},t) 
=
-\int^{t}_{0}d\tau \Gamma ({\bf k},\tau)\delta \sigma({\bf k},t-\tau) 
+ \xi ({\bf k},t) \label{eqn:TC}.
\end{eqnarray}
Here, the memory function $\Gamma({\bf k},\tau)$ is given by the inverse 
Laplace transformation of $\Gamma^{L}({\bf k},s)$, defined by 
\begin{eqnarray}
\Gamma^{L}({\bf k},s) 
= - s\dot{\chi}^{L}_{s}({\bf k})
/(\chi_{0}({\bf k})+\dot{\chi}^L_s({\bf k}))\cdot
(1-2g\beta\chi_{0}({\bf k})), \label{eqn:gamma_L}
\end{eqnarray}
where 
the index $L$ means the Laplace transformation, 
$\dot{\chi}^L_s({\bf k}) = \int^{\infty}_{0}dt e^{-st} d\chi_{t}({\bf k})/dt$ 
with 
\begin{eqnarray}
\chi_t({\bf k})
&=& \frac{N_{c}N_{f}}{\beta V}
\sum_{\bf p}
\{
(1-2n(E_{\bf p}))\frac{E_{\bf p}E_{\bf p+k}+{\bf p(p+k)} }
{E_{\bf p}E_{\bf p+k}(E_{\bf p} + E_{\bf p+k})}\nonumber \\
&&\times 
(e^{i(E_{\bf p}+E_{\bf p+k})t} + e^{-i(E_{\bf p} + E_{\bf p+k})t}) \nonumber \\
&& + (n(E_{\bf p+k}) - n(E_{\bf p})) 
\frac{E_{\bf p}E_{\bf p+k} - {\bf p(p+k)} }
{E_{\bf p}E_{\bf p+k}(E_{\bf p} - E_{\bf p+k})}\nonumber \\
&&\times (e^{i(E_{\bf p} - E_{\bf p+k})t} + e^{-i(E_{\bf p} - E_{\bf p+k})t})
\}, 
\end{eqnarray}
Here, $n(E) = [\exp (\beta E)+1 ]^{-1}$, 
$E_{\bf k} = |{\bf k}|$, $N_{c}=3$ and $N_{f}=2$.
It should be noted that there is no chiral condensate above $T_{c}$.
In the calculation of the memory term, 
we calculated only the term corresponding to a ring diagram 
in the scalar channel 
and employed the technique proposed in Ref. \cite{ref:Sawada}.

The memory function 
$\Gamma^{L}({\bf 0},s)$ vanishes at $T_{c}$.
This is shown by using the self-consistency condition of 
the chiral condensate calculated in the mean field approximation,
\begin{eqnarray}
1-2g\beta \chi_{0}({\bf 0})|_{T=T_{c}} = 0,\label{eqn:SCC}
\end{eqnarray}
where $T_{c}=185$ MeV.
From Eqs. (\ref{eqn:gamma_L}) and (\ref{eqn:SCC}), 
one can see that $\Gamma^{L}({\bf 0},s)$ vanishes at $T_{c}$.
The small $\Gamma^{L}({\bf 0},s)$ near $T_{c}$ means 
the slowing down of the relaxation, that is nothing but CSD.
If we take account of the terms beyond the ring diagram approximation in the 
calculation of the memory function, 
the critical temperature deviates from that in the mean field approximation.

\begin{figure}[t]
\begin{center}\leavevmode
\epsfxsize=6cm
\epsfbox{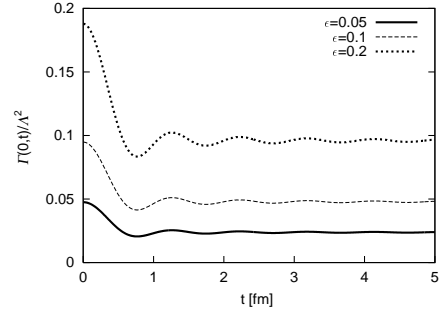} 
\caption{
The time evolution of $\Gamma({\bf 0},t)$ in the unit of $\Lambda^2$.
The solid, dashed and dotted lines are the $\Gamma({\bf 0},t)/\Lambda^2$ 
for the temperatures 
$\epsilon=(T-T_{c})/T_{c}=0.05, 0.1$ and $0.2$, respectively.
}
\end{center} 
\end{figure}

The TC equation (\ref{eqn:TC}) is still inadequate to be interpreted as 
a Langevin equation because the memory function $\Gamma({\bf k},t)$ 
converges to a finite value at $t \rightarrow \infty$, as is shown in Fig.~1.
As a matter of fact, 
it is usually expected that the memory function tends to zero 
with the shorter time scale than the macroscopic one, 
because we can prove that the memory function is given by the time 
correlation of the noise.
This is called the 2nd F-D theorem\cite{ref:Mori}.
The anomalous behavior of the memory function gives rise to 
the oscillation in the time evolution of $\delta \sigma({\bf k},t)$, 
as we will see later in Fig.~2.
This anomaly may arise because 
the MPO defined in Eq. (\ref{eqn:MPO})
is incomplete such that the noise still incorporates 
gross variables other than $\delta \sigma({\bf x})$.
As an example, let us consider the Langevin equation 
with two gross variables $x$ and $p$,\cite{ref:Zwanzig}
\begin{eqnarray}
\frac{d}{dt}x(t) &=& p(t), \\
\frac{d}{dt}p(t) &=& - \omega^2 x(t) - \int^{t}_{0}ds \Xi(t-s)p(s) + f(t),
\end{eqnarray}
where $\Xi(t)$ is the memory function 
given by the time correlation of the noise $f(t)$ and 
has no long time correlation.
Solving the first equation and substituting into the second one, 
we have
\begin{eqnarray}
\frac{d}{dt}p(t) = - \int^{t}_{0}ds (\omega^2 + \Xi(t-s))p(s) + f(t),
\end{eqnarray}
where we set the initial condition $x(0)=0$.
Although the equation has only one gross variable $p$, 
the memory function $\omega^2 + \Xi(t)$ has 
the long time correlation due to the oscillation.
In this case, the true memory term given by the 2nd F-D theorem is 
not $\omega^2 + \Xi(t)$ but $\Xi(t)$.

Thus, to define a {\it renormalized} memory function without 
long time correlation, 
we separate the oscillation effect from the memory function.
By carrying out the Fourier transformation, 
one can easily recognize 
that the correction to the frequency shift and the damping are given by 
the imaginary part and the real part of the Fourier transform of 
the memory function, 
respectively\cite{ref:Kubo-Stat}.
Thus, the Langevin equation with the renormalized 
memory function $\Phi({\bf k},t)$ is expressed as,
\begin{eqnarray}
\frac{d}{dt} \delta \sigma({\bf k},t) 
&=& -\int^{t}_{0}d\tau \Omega^2_{\bf k} (t-\tau) \delta \sigma({\bf k},\tau) 
\nonumber \\
&&\hspace*{-1cm} -\int^{t}_{0}d\tau \Phi 
({\bf k},\tau)\delta \sigma({\bf k},t-\tau) 
+ \xi ({\bf k},t), \label{eqn:QLE2}
\end{eqnarray}
where 
$\Omega^{2}_{\bf k}(t) 
= i\int d\omega {\rm Im}[\Gamma^{L} ({\bf k},-i\omega+\epsilon)] 
e^{-i\omega t}/2\pi$ 
and 
$\Phi({\bf k},t) 
= \int d\omega {\rm Re}[\Gamma^{L}({\bf k},-i\omega+\epsilon)] 
e^{-i\omega t}/2\pi$.
This memory function does not have a long time correlation any more.
Here, we artificially removed the oscillation effect from the memory term.
However, a similar procedure should be automatically 
implemented by using the MPO defined 
by a complete set of gross variables.

Now, we can define the noise with short time correlation.
The definition of the noise and the 2nd F-D theorem lead to the 
following correlation properties:
\begin{eqnarray}
( \xi({\bf k},t), \delta \sigma({\bf k}',0) )
&=& \langle \xi({\bf k},t) \rangle = 0, \\
(\xi({\bf k},t),
\xi^{\dagger}({\bf k'},t'))
&=& V\delta^{(3)}_{\bf k,k'} \Phi({\bf k},t-t')\chi_{0}({\bf k}),
\end{eqnarray}
where $\langle O \rangle = {\rm Tr}[\rho~O]$ and $V$ 
is the volume of the system.
%$(X,Y)$ is the canonical correlation, again.
Here, we assume that the noise has 
the translational invariance in space and time.
The first correlation indicates that the noise 
does not include components that vary with the gross time scale 
and it is possible to regard it as 
a random field for $\delta \sigma({\bf k},t)$.
The second correlation characterizes the noise as a random field.
In the following, 
we solve Eq.~(\ref{eqn:QLE2}) as a classical equation with a random noise.
For this, we introduce the classical noise that reproduces the correlation 
properties defined above;
\begin{eqnarray}
\ll \xi({\bf k},t)\delta \sigma({\bf k}',0) \gg &=& \ll \xi({\bf k},t) \gg
= 0, \\
\ll \xi({\bf k},t)\xi^{*}({\bf k'},t') \gg 
&=& (\xi({\bf k},t),
\xi^{\dagger}({\bf k'},t')), \label{eqn:flu-corre2}
\end{eqnarray}
where $\ll~\gg$ means the average for noise with a suitable 
stochastic weight.
It is worth emphasising that 
the correlations determined here coincide with 
the condition of themalization for a Langevin equation\cite{ref:Kubo-Stat}.
In this sense, the system described by the Langevin equation 
approaches a thermal equilibrium state with time.

The averaged time evolution of the critical fluctuation at vanishing momentum 
is shown in Fig.~2.
One can see that the nonequilibrium fluctuation 
relaxes with oscillation and finally converges zero.
%That is, the relaxation of the critical fluctuation is underdamping.
This indicates that the critical dynamics of the chiral transition 
may not be described 
by a diffusion equation like the time dependent Ginzburg-Landau (TDGL) 
equation.
The rate of the relaxation becomes slower as the temperature 
approaches $T_{c}$ because of CSD.
The relaxation time is characterized by $\tau_{rt} = 2/\gamma_{\bf 0}$ where 
$2\gamma_{\bf k} = \int^{\infty}_{0}ds \Phi({\bf k},s)$.
At $\epsilon =(T-T_{c})/T_{C} = 0.2$, $\tau_{rt}$ is about $10$ fm, 
that is the same order as 
the expected life time of QGP.
Thus, we cannot ignore the fluctuation of the order parameter 
at a temperature lower than $222$ MeV.
On the other hand, at higher temperature, 
the fluctuation relaxes firstly and other gross variables 
become important.

\begin{figure}[t]
\begin{center}\leavevmode
\epsfxsize=6cm
\epsfbox{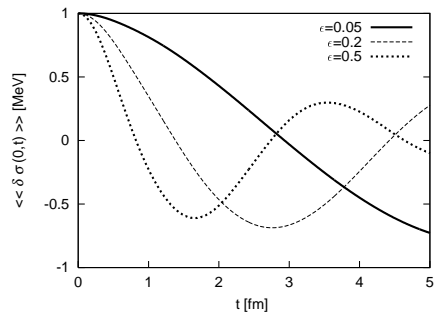} 
\caption{
The time evolution of $\delta \sigma({\bf 0},t)$.
The solid, dashed and dotted lines are the $\delta \sigma({\bf 0},t)$
for the temperatures $\epsilon=0.05, 0.2$ and $0.5$, respectively.
}
\end{center} 
\end{figure}

As we have shown so far, the Langevin equation reveals CSD 
and thermalization.
This means that the Langevin equation correctly describes the behavior 
near $T_{c}$.
At the same time, as was pointed out in Ref. \cite{ref:HK}, 
a soft mode appears above $T_{c}$ in the correlation function of 
the fluctuation of the order parameter.
To investigate the correlation function, 
we define the power spectrum as 
\begin{eqnarray}
I({\bf k},\omega) = 
\lim_{T,V \rightarrow \infty}
\frac{1}{TV}
\ll |\delta \sigma ({\bf k},\omega)|^2 \gg,
\end{eqnarray}
where $T$ is the time when we observe the system, and 
\begin{eqnarray}
\ll |\delta \sigma({\bf k},\omega)|^2 \gg 
&=& TV 
\frac{{\rm Re}~[\omega^2 \Phi({\bf k},\omega)\chi_{0}({\bf k})]}
{|-\omega^2 + \Omega_{\bf k}^2 -i\omega \Phi ({\bf k},\omega)|^2}.
\end{eqnarray}
The Wiener-Khinchin theorem tells us that the correlation function is 
given by the power spectrum 
\begin{eqnarray}
C({\bf x},t)
&=& \lim_{t'\rightarrow \infty}
\ll \delta \sigma({\bf x+x'},t+t') \delta \sigma({\bf x'},t') \gg 
\nonumber \\
&=& \int^{\infty}_{-\infty}\frac{d\omega d^3 {\bf k}}{(2\pi)^4}
I({\bf k},\omega) e^{-i\omega t}e^{i{\bf k,x}}.
\end{eqnarray}
The temperature dependence of the power spectrum at vanishing momentum 
is shown in Fig.~3.
We can see that the peak moving toward origin 
becomes prominent as the temperature is lowered toward $T_{c}$.
The power spectrum characterizes the space-time correlation 
in energy-momentum space 
and hence can be interpreted as the spectral function 
in the thermal Green's function.
Then, the peak with narrow width 
reveals the existence of a collective mode whose 
energy tends to vanish as the temperature approaches $T_{c}$.
Such a mode is called a soft mode.
The temperature dependence of the power spectrum is consistent with 
the previous result, where the spectral function is 
calculated in the linear response theory\cite{ref:HK}.

\begin{figure}[t]
\begin{center}\leavevmode
\epsfxsize=6cm
\epsfbox{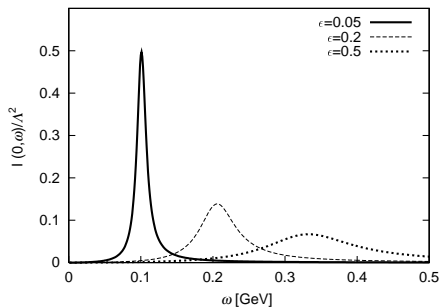} 
\caption{
The temperature dependence of $I({\bf 0},\omega)$ in the unit of $\Lambda^2$.
The solid, dashed and dotted lines are the $I({\bf 0},\omega)/\Lambda^2$
for the temperatures $\epsilon=0.05, 0.2$ and $0.5$, respectively.
}
\end{center} 
\end{figure}

We have derived the linear Langevin equation 
that describes the dynamics of the chiral phase transition in the NJL model 
using the projection operator method.
The equation reveals CSD and shows thermalization.
The order parameter relaxes exhibiting oscillation.
This means that a simple diffusion-type equation like 
the TDGL equation may be 
inadequate to describe the dynamics of the chiral transition.
The power spectrum was also calculated 
using the Langevin equation 
and we found that there exists a soft mode.
As a result, we can conclude that the Langevin equation 
fulfils the requirements near $T_{c}$: 
CSD, thermalization and soft mode.

We have discussed here only 
the linear Langevin equation and ignored the nonlinear effect.
However, the mode coupling theory makes it clear that near critical points 
the nonlinear effect becomes important because of the large correlation 
length and leads to the deviation from the van Hove theory 
in calculating dynamical critical exponent\cite{ref:Kawasaki}.
In order to take nonlinear terms into account, 
we must choose them as gross variables in defining the MPO.
Another intriguing subject 
is to apply this formulation to finite density 
and the color superconducting phase transition.
These are future projects.

The authors thanks M.~Asakawa, A.~Onuki, T.~Hirano, A.~Muronga, 
D.~Rischke and I.~Sawada for fruitful discussions.
T.K. acknowledges a fellowship from the Alexander von Humboldt 
Foundation.

\end{document}